\documentclass[aps,prl,reprint,superscriptaddress,showpacs,longbibliography,floatfix,footnoteinbib]{revtex4-1}
\usepackage[latin9]{inputenc}
\usepackage{xcolor}
\usepackage{bm}
\usepackage{amsmath}
\usepackage{amssymb}
\usepackage{graphicx}
\usepackage{wasysym}
\PassOptionsToPackage{normalem}{ulem}
\usepackage{ulem}
\usepackage[unicode=true,
 bookmarks=false,
 breaklinks=false,pdfborder={0 0 1},backref=false,colorlinks=true]
 {hyperref}
\hypersetup{pdftitle={Title},
 plainpages=false,pdfpagelabels,linkcolor=red,urlcolor=blue,citecolor=blue,pdfdisplaydoctitle=true,pdfduplex=DuplexFlipLongEdge}

\makeatletter

\providecommand{\tabularnewline}{\\}
\providecolor{lyxadded}{rgb}{0,0,1}
\providecolor{lyxdeleted}{rgb}{1,0,0}
\DeclareRobustCommand{\lyxadded}[3]{{\texorpdfstring{\color{lyxadded}{}}{}#3}}

\DeclareRobustCommand{\lyxsout}[1]{\ifx\\#1\else\sout{#1}\fi}
\usepackage{tikz}
\usetikzlibrary{calc}

\usepackage{units}\usepackage{wasysym}

\usepackage{bm}
\usepackage{braket}

\makeatother

\begin{document}
\title{Delocalization of topological surface states by diagonal disorder
in nodal loop semimetals}
\author{Jo\~ao S. Silva}
\affiliation{Centro de F\'{\i}sica das Universidades do Minho e Porto, Departamento
de Física e Astronomia, Faculdade de Ciências, Universidade do Porto,
4169-007 Porto, Portugal}
\author{Miguel A. N. Ara\'ujo}
\affiliation{CeFEMA, Instituto Superior T\'ecnico, Universidade de Lisboa, Av. Rovisco
Pais, 1049-001 Lisboa, Portugal}
\affiliation{Departamento de F\'{\i}sica, Universidade de \'Evora, P-7000-671, Évora,
Portugal}
\affiliation{Beijing Computational Science Research Center, Beijing 100084, China}
\author{Miguel Gon\c{c}alves}
\affiliation{CeFEMA, Instituto Superior T\'ecnico, Universidade de Lisboa, Av. Rovisco
Pais, 1049-001 Lisboa, Portugal}
\author{Pedro Ribeiro}
\affiliation{CeFEMA, Instituto Superior T\'ecnico, Universidade de Lisboa, Av. Rovisco
Pais, 1049-001 Lisboa, Portugal}
\affiliation{Beijing Computational Science Research Center, Beijing 100084, China}
\author{Eduardo V. Castro}
\affiliation{Centro de F\'{\i}sica das Universidades do Minho e Porto, Departamento
de Física e Astronomia, Faculdade de Ciências, Universidade do Porto,
4169-007 Porto, Portugal}
\affiliation{Beijing Computational Science Research Center, Beijing 100084, China}
\begin{abstract}
The effect of Anderson diagonal disorder on the topological surface
(``drumhead'') states of a Weyl nodal loop semimetal is addressed.
Since diagonal disorder breaks chiral symmetry, a winding number cannot
be defined. 
Seen as a perturbation, the weak random potential mixes the clean
exponentially localized drumhead states of the semimetal, thereby
producing two effects: \textit{(i)} the algebraic decay of the surface
states into the bulk; \textit{(ii)} a broadening of the low energy
density of surface states of the open system due to degeneracy lifting.
This behavior persists with increasing disorder, up to the bulk semimetal-to-metal
transition at the critical disorder $W_{c}$. 
Above $W_{c}$, the surface states hybridize
with bulk states and become extended into the bulk. 
\end{abstract}
\maketitle

\section{Introduction}

The most appealing property of topological matter is the robustness
of certain material properties to perturbations. Among such properties,
the creation of robust
localized states at the edge or surface of a sample is probably the
most striking.
Rooted in the bulk-edge correspondence, these topologically protected
edge states survive weak disorder, which makes them appealing also
from the point of view of applications.
Three dimensional
(3D) topological
insulators, with their two-dimensional (2D) Dirac-fermions at the surface, stood out
as an important class of topological materials \citep{HKrmp10,QZrmp11}
whose stability with respect to interactions and disorder is by now
fairly well established \citep{ChiuRMP2016,Rachel2018}. Gapless systems
can, however, also support non-trivial momentum-space topology
and robust, topologically protected, surface states.
 Among these
are the Weyl nodal loop (WNL) semimetals, for which the valence and
conduction bands linearly touch along one-dimensional (1D) loops in
3D momentum space \citep{Armitage2018}. 
Their recent theoretical prediction \citep{Kim2015,Weng2015,PhysRevLett.115.026403,Chen2015}
and experimental discovery \citep{Xie2015,Bian2016} triggered intense
experimental \citep{Schoop2016,Okamoto2016,Xu2017a,Lou2018,PhysRevB.99.241102,Qiu2019,Sims2019}
and theoretical interest \citep{Rhim2015,Fang2015,Huang2016,Chan2016,Lu2017,Xu2017,
Du2017,balents2017nodalLoop,Oroszlany2018,Martin-Ruiz2018,ShuChen2018,Lau2019,Ezawa2019,linhu2016,engineering2017}. 

The WNL's topological nature manifests itself by the presence of localized
(``drumhead'') states \citep{Araujobook2021,Burkov2011,Weng2015,Chan2016,Zhang2016}.
In the thermodynamic limit, the surface states have zero energy and
produce a delta-function contribution to the
bulk density of states (DOS) 
of the open system, $\rho_{{\rm edge}}(E)\propto A_{\newmoon}\delta(E)$,
where $A_{\newmoon}$ denotes the $\bm{k}$-space
area of the nodal loop projected onto the surface.
This has to be contrasted with the bulk density of states: since the
Fermi surface is reduced to a 1D nodal line, the bulk DOS, $\rho_{{\rm bulk}}(E)$,
vanishes linearly for low energies, i.e. $\rho_{{\rm bulk}}(E)\propto\left|E\right|$.
Drumhead states have already been observed experimentally through
angle-resolved photoemission spectroscopy (ARPES), transport measurements
and de Haas--van Alphen quantum oscillations \cite{Schoop2016,Sims2019,Nakamura2019,Belopolski2019,observation_drumhead,Stuart2022}.

The effect of
static disorder on the bulk properties has been addressed recently
\citep{Goncalves2020}.
Within an Anderson model of box-distributed disorder, a phase transition
was found from a bulk low disorder multifractal semimetallic (SM)
phase, where the momentum-space wave-function has multifractal structure,
to a single-fractal diffusive metallic (M) phase. This SM/M transition
takes place at a finite disorder value, $W_{c}$ \footnote{For Gaussian distributed
disorder, rare regions effects yield, instead, an avoided quantum
critical point \citep{Pixley2016}}.

For a WNL, the fate of the topological drumhead states under finite
disorder is yet unknown. To our knowledge, only the effect of an incommensurate
potential on the drumhead states of a nodal link semimetal has been
addressed \citep{ShuChen2018}. However, that work assumed the potential
to depend only on
one spatial coordinate, while translational invariance in the perpendicular
plane was preserved \footnote{We also mention here that the dissolution 
of Fermi arcs by unidirectional disorder  in a dirty Weyl point semimetal
has also been addressed \citep{Slager2017}}. 
A study of three-dimensional disorder effects
on the surface states of a WNL is in order, then. 

In this work, we unveil the fate of the topological surface states
of the open WNL system under Anderson short ranged diagonal disorder.
Because the latter breaks chiral symmetry, a winding number cannot
be defined. However, the surface states can be detected by studying
(i) the density of surface states (DOSS), $\Delta\rho(E)$, defined
as the change in the DOS when a surface is created in the direction
perpendicular to the nodal loop plane; and (ii) the localization properties
of the topological surface states with increasing disorder.
The latter, however weak, always produces the broadening of the DOSS. 
Yet, the total number of surface states is found to decrease monotonically up to very strong disorder 
in an approximately exponential form.
Concomitantly, the finite energy surface states become delocalized upon hybridization with bulk states.
The zero energy surface states decay algebraically into the bulk while remaining square integrable for weak disorder,
and become extended at the bulk SM/M transition.

The rest of the paper is organized as follows.
In Sec.~\ref{sec:Model-and-Methods}
we present the model and provide details on the methods used in this
work. 
The results are given in Sec.~\ref{sec:Results}.
In Sec~\ref{sec:Discussion} a final discussion is provided. 

\section{Model and Methods}
\label{sec:Model-and-Methods}

We study a two-band model of a WNL system on a cubic lattice with
diagonal disorder \citep{Goncalves2020}, 
\begin{equation}
H=\sum_{\bm{k}}c_{\bm{k}}^{\dagger}H_{\bm{k}}c_{\bm{k}}+\sum_{\bm{r}}c_{\bm{r}}^{\dagger}V_{\bm{r}}(W)c_{\bm{r}}\,.\label{eq:total_H}
\end{equation}
The first term describes a clean WNL semimetal,
with $\bm{k}$ a 3D Bloch vector, $H_{\bm{k}}=(t_{x}\cos k_{x}+t_{y}\cos k_{y}+\cos k_{z}-m)\tau_{x}+t_{2}\sin k_{z}\tau_{y}$,
with $\tau_{x},\tau_{y}$ Pauli matrices acting on the orbital pseudo-spin
indices $\alpha=1,2$, and $c_{\bm{k}}^{\dagger}=(\begin{array}{cc}
c_{\bm{k},1}^{\dagger} & c_{\bm{k},2}^{\dagger}\end{array})$, where $c_{\bm{k},\alpha}^{\dagger}$
creates an electron with Bloch momentum $\boldsymbol{k}$ in the sublattice
spanned by $\alpha$ orbitals. 
The clean nodal loop system, $H_{\bm{k}}$,
is chiral symmetric as it anticommutes
with the operator $\tau_z$.
The second term is the disorder potential,
where $\boldsymbol{r}$ denotes a lattice site and $V_{\bm{r}}(W)=\text{diag}(v_{\bm{r}1},v_{\bm{r}2})$,
with random variables $v_{\bm{r}\alpha}\in[-W/2,W/2]$, where $W$
corresponds to the disorder strength. We ensure that $v_{\bm{r}\alpha}$
averages to zero in sublattice $\alpha=1,2$ for each disorder realization.
The results presented hereafter are for $t_{x}=1.1$, $t_{y}=0.9$,
$m=2.12$ and $t_{2}=0.8$. The hopping anisotropy chosen breaks unwanted
degeneracies and ensures the system is generic within this class.
This parameter choice yields a single nodal line, in the $k_{z}=0$
plane, given by: 
\begin{equation}
t_{x}\cos k_{x}+t_{y}\cos k_{y}+1-m=0\,.\label{nodalline}
\end{equation}
(see Ref \citep{Goncalves2020}).

Because the diagonal disorder breaks chiral symmetry, a winding number
cannot be defined. 
However, the surface states can be detected by
studying the DOSS, $\Delta\rho(E)$, defined as 
 $\Delta\rho(E)\equiv\rho_{{\rm open}}(E)-\rho_{{\rm bulk}}(E)$.
Here, $\rho_{{\rm bulk}}$ denotes the DOS calculated for periodic
boundary conditions (PBC), and $\rho_{{\rm open}}$ denotes the DOS
calculated using open boundary conditions (OPB) along the $z$ direction
(perpendicular to the nodal loop plane).
 If one would only consider an open system,
identifying surface states would require knowing the local DOS of the
eigenstates to see which are localized at the surface.
This is why we instead compare an open with a closed system and compute the DOSS: 
any change in the DOS must be a surface effect. 
 To compute the DOS we use
the kernel polynomial method (KPM) with an expansion in Chebyshev
polynomials to order $N_{m}$ \cite{Joao,Joao2020},
reaching system sizes containing up to $L=100$ unit cells in each
direction.

Exact diagonalization (ED) using the Lanczos method allows us to study
the localization properties of the topological surface states. The
surface states' localizaton along the $z$ direction is revealed by
an inverse participation ratio defined for the $z$ direction in sublattice
$\alpha$ as 
\begin{equation}
{\rm IPR}_{z}^{\alpha}=\frac{\sum_{z}\Psi^{4}(z,\alpha)}{\left[\ \sum_{z}\Psi^{2}(z,\alpha)\ \right]^{2}}\,,\label{iprzalfadef}
\end{equation}
with
\begin{equation}
\Psi^{2}(z,\alpha)=\sum_{x,y}|\psi(x,y,z;\alpha)|^{2}\,,\label{probz}
\end{equation}
where $\psi(x,y,z;\alpha)$ is the eigenstate amplitude in the lattice
cell at $(x,y,z)$ and orbital $\alpha$.

\section{Results}
\label{sec:Results}

\subsection{Clean system}

For a better understanding of the effect of diagonal disorder on drumhead surface states, 
we first review the clean system. 
Let us write the momentum as $\bm{k}=(\bm{k}_{\parallel},k_{z})$,
where $\bm{k}_{\parallel}=(k_{x},k_{y})$ is the momentum component
parallel to the surface. A winding number for each $\bm{k}_{\parallel}$
can be defined \citep{Araujobook2021,LiMiguel2017}, 
\begin{equation}
\begin{aligned}
\nu(\bm{k}_{\parallel})
=\frac{1}{2\pi}\int_{-\pi}^{\pi}dk_{z}\langle\psi(\bm{k}_{\parallel})|\partial_{k_{z}}|\psi(\bm{k}_{\parallel})\rangle\\
=\frac{1}{2\pi}\int_{-\pi}^{\pi}\frac{\partial\log H_{12}(\bm{k}_{\parallel},k_{z})}{\partial k_{z}}dk_{z}\,,
\end{aligned}
\label{winding}
\end{equation}
where $H_{12}=t_{x}\cos k_{x}+t_{y}\cos k_{y}+\cos k_{z}-m-it_{2}\sin k_{z}$
is the off-diagonal matrix element of the clean WNL Bloch Hamiltonian. 
Appealing to dimensional reduction, one can take $\bm{k}_{\parallel}$
as a label for a topological chain \citep{Pershoguba2012,Shockley}
along the $z$ direction.

Equation ~(\ref{winding}) yields $\nu(\bm{k}_{\parallel})=1$ or 
$\nu(\bm{k}_{\parallel})=0$ if $\bm{k}_{\parallel}$
is inside or outside the nodal loop, respectively. This implies that for each $\bm{k}_{\parallel}$
inside the loop, there is a zero-energy surface state 
$\psi_{\bm{k}_{\parallel}}(\bm{r},\alpha=1)=e^{i(k_{x}x+k_{y}y)}\phi_{\bm{k}_{\parallel}}(z)$,
occupying sublattice $\alpha=1$, where $\phi_{\bm{k}_{\parallel}}(z)$
decays exponentially from the surface at $z=1$ of a semi-infinite system, $z\geq1$. 
For a finite system
with linear size $L$, a similar surface state exists on the opposite
surface, $z=L$, occupying sublattice $\alpha=2$. In the thermodynamic
limit, such states have zero energy. For finite $L$, the small hybridization
between drumhead states in opposite surfaces lifts their degeneracy.
As $\bm{k}_{\parallel}$ approaches the nodal line, the decay length
of $\phi_{\bm{k}_{\parallel}}(z)$ diverges, thereby increasing the
finite-size hybridization.

The number of surface states ($p$) created
by opening a cubic system at $z=1$ and $z=L$ is just twice the number
of $\bm{k}_{\parallel}$ points inside the nodal line in Eq.$\,$(\ref{nodalline}),
and is thus proportional to the loop area
($A_{\newmoon}$) in $\bm{k}_{\parallel}$-space times $L^{2}$,
\begin{equation}
p=\Lambda L^{2},\label{eq:numbSS}
\end{equation}
where we have defined $\Lambda\equiv A_{\CIRCLE}/2\pi^{2}$. 
Since the DOS is defined per unit volume,
the clean system's DOSS in the $L\rightarrow\infty$ limit is
\begin{equation}
\Delta\rho(E)=\Lambda L^{-1}\delta(E)+f_{reg}(E)\,,\label{dossL}
\end{equation}
where the $\delta$-function accounts for the topological zero-energy
surface states and $f_{reg}(E)$ is a regular function. The number
of zero-energy drumhead states 
may then be obtained from $\Delta\rho(E)$, 
\begin{equation}
p=L^{3}\int_{0^{-}}^{0^{+}}\Delta\rho(E)\ dE=\Lambda L^{2}\,.\label{alfa}
\end{equation}
Note that for a given system size, the creation of edge states removes
states from the bulk  without changing the total number of eigenstates, 
which is $2L^3$,
therefore, 
\mbox{$\int_{-\infty}^{\infty}\Delta\rho(E)dE=0$}.

\begin{figure}
\noindent \centering{}\includegraphics[width=0.95\columnwidth]{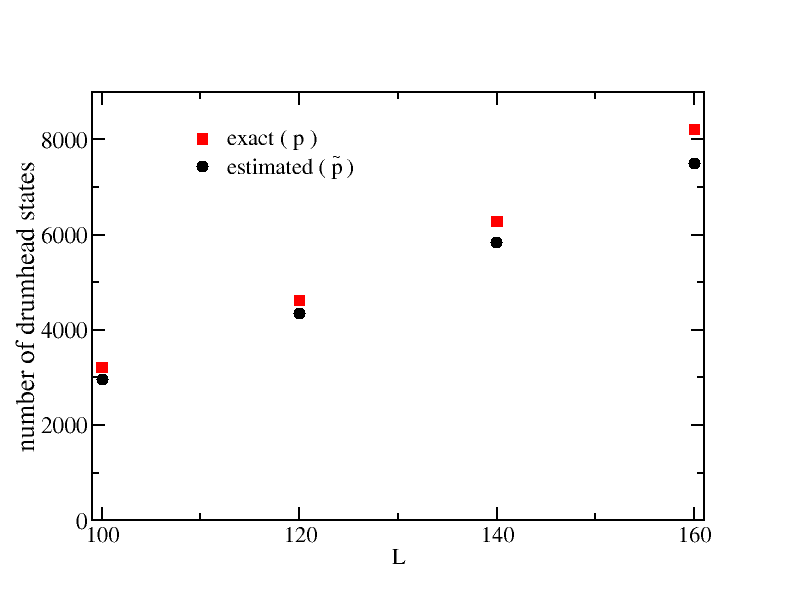}
\caption{Clean system's exact ($p$, red) and estimated ($\tilde{p}$, black)
number of surface states for different system sizes containing $L^{3}$
	unit cells. $N_m=2^{10}$.}
\label{number_states_fig}
\end{figure}

Although the clean WNL is analytically tractable, it can serve as
a test bed for the DOS calculation through the numerical KPM method. 
The finiteness of $L$ necessarily causes some
broadening of the delta function due to the hybridization explained
above. Therefore, one must integrate $\Delta\rho(E)$ over an energy
interval in order to obtain the number of edge states, $p$. The best
choice is to define an energy window, $E_{w}$, such that $\Delta\rho(E)>0$
for $|E|<E_{w}$ and numerically estimate the number of surface states
as 
\begin{equation}
\tilde{p}=L^{3}\int_{-E_{w}}^{E_{w}}\Delta\rho(E)\ dE\,,\label{tildep}
\end{equation}
which can be compared to the exact value, $p$.
In Tab.~\ref{ptable} we collect a few examples of $p$ and $\tilde{p}$.
An analogous comparison between $p$ and $\tilde{p}$ as functions
of system size $L^{3}$ is shown in Fig.~\ref{number_states_fig}.
The integral in Eq.~\eqref{tildep} nearly captures the exact number
of edge states, though a finite difference persists even for the larger
sizes.
\begin{table}[htp]
\begin{centering}
\begin{tabular}{ccc}
\hline 
~~~~~$L_{x}\times L_{y}\times L_{z}$ & ~~~~$p$
~~~~ & $\tilde{p}$ ~~~~~~~\tabularnewline
\hline 
\hline 
80$\times$80$\times$80 & 2050  & 1965 (96\%) \tabularnewline
80$\times$ 80$\times$160  & 2050  & 1970 (96\%)\tabularnewline
80$\times$80$\times$200  & 2050  & 1974 (96\%)\tabularnewline
80$\times$80$\times$240  & 2050  & 1978 (96\%)\tabularnewline
100$\times$100$\times$100  & 3214  & 3104 (97\%)\tabularnewline
\hline 
 &  & \tabularnewline
\end{tabular}
\par\end{centering}
\caption{Clean system's exact ($p$) and estimated ($\tilde{p}$) number of
surface states. $L_{\mu}$ corresponds to the number of unit cells
in direction $\mu=x,y,z$. In brackets, $\tilde{p}/p$ is given in
percentage. The number of polynomials is $N_{m}=\{2^{11},2^{12}\}$ .}
\label{ptable} 
\end{table}

As stated before, the difference $p-\tilde{p}$ is due to hybridization
between edge states on opposite sides of the sample when $\bm{k}_{\parallel}$
lies close to the nodal line in Eq.$\,$(\ref{nodalline}). The energy
splitting displaces some states to outside of the energy window $E_{w}$. 
We expect that the hybridization effect
is reduced when the system's size along $z$ is increased, allowing
more states to get captured by Eq.(\ref{tildep},
but the effect is relatively small, as Tab.~\ref{ptable} shows.
Because the localization length diverges near the nodal line,
the difference  $p-\tilde{p}$ remains proportional to the nodal line perimeter,
so it is expected to scale with $L$.

\subsection{Disordered system}

\subsubsection{Spectral properties}

Anderson disorder, however weak, has two effects:
\emph{(i)} chiral symmetry breaking, and \emph{(ii)} hybridization
between the clean system's drumhead states.

\textit{(i)} Because of chiral symmetry breaking, 
a winding number cannot be defined. The winding number calculated
from the eigenstates (Zak phase) in Eq.~(\ref{winding}) is not
quantized for a system without chiral symmetry and therefore it cannot
measure the number of edge states. This holds true for the real space
formulation of the Zak phase - the correct formulation for systems
with broken translational invariance - where $\bm{k}$ can be replaced
by phase twists $\bm{\theta}$ in Eq.$\,$(\ref{winding}) \cite{vanderbilt_2018}.

\begin{figure}[htp]
\centering{}\includegraphics[width=0.95\columnwidth]{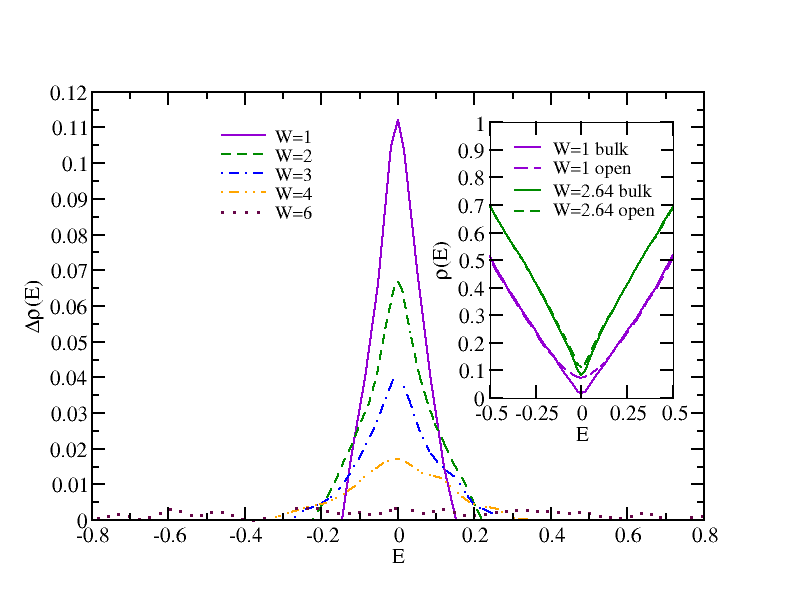}
\caption{Averaged DOSS. $\langle\Delta\rho(\epsilon)\rangle>0$ defines the energy window
$\left[-E_{w},E_w\right]$, which is found to grow with increasing $W$. Results averaged
over 20 disorder realizations and $L=100$.
Inset: $\rho_{bulk}(E)$ and $\rho_{open}(E)$ for two values of $W$.} \label{DDOS_E_W}
\end{figure}

\textit{(ii)} Due to hybridization between all the clean system's drumhead states, 
their energies are shifted, causing a disorder broadening of the DOSS
near zero energy. An example of the DOSS
broadening around $E=0$, averaged over disorder realizations, is
shown in Fig.~\ref{DDOS_E_W}, 
where the smallest disorder considered is $W=1$, much smaller than
the bandwidth of the clean system.

Under increasing disorder strength, $W$, the energy window, $E_{w}$, 
of Eq.~\eqref{tildep} also increases. This is seen in Fig.~\ref{DDOS_E_W}, where a notable $\Delta\rho(E)>0$
for an increasing energy window around $E=0$ is present even for
$W=4$. 
These results suggest that some sort of edge states, reminiscent of the
topological drumhead states, persist at higher disorder, even though
topological protection is not at work due to chiral symmetry breaking.
In Sec.~\ref{subsec:Localization-properties} we provide details
on the localization properties of these edge states. 

It should be noted that for $W\gtrsim2.6$, the bulk develops a nonzero
$\rho_{{\rm bulk}}(0)$, entering the metallic diffusive phase \citep{Goncalves2020}.
For $W>6$ the DOS around zero energy no longer resembles 
that of a semimetalic WNL, which is characterized by a linearly vanishing DOS.
In our model, the linear dependence occurs in an energy scale $|E|\lesssim0.8$ \citep{Goncalves2020}.
For $W>6$ we obtain $E_{w}\gtrsim0.8$, therefore, larger than the
energy scale characterizing a WNL in
the semimetal phase. 
This means that $W=6$ is a very strong disorder. Incidentally,
we note that at $E=0$, Anderson localization in this model occurs
for a much stronger disorder of $W_{A}\approx11$ \citep{Goncalves2020}.

\begin{figure}[htp]
\centering{}\includegraphics[width=0.95\columnwidth]{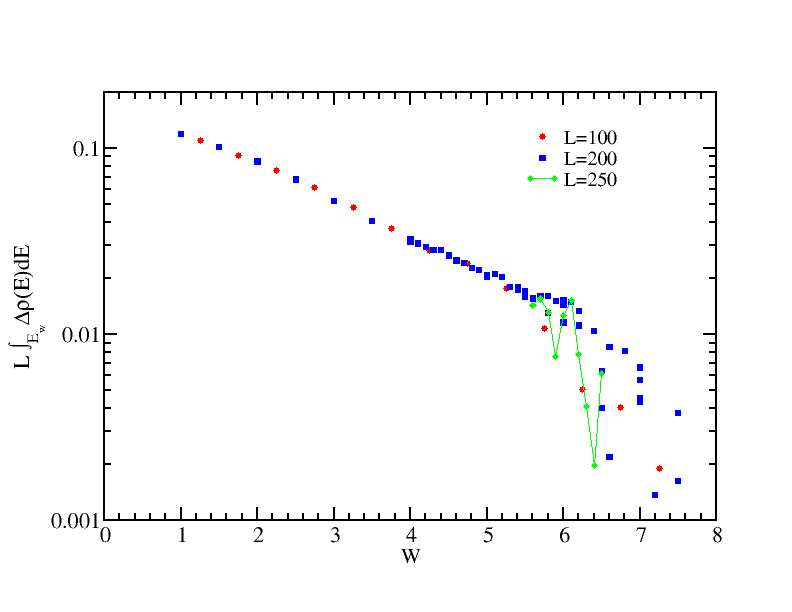}
\caption{Logarithmic plot of $L\int\Delta\rho=\tilde{p}/L^2$ for different $L$.
A reasonable collapse is observed for $W<6$, indicating $L^{-1}$ scaling
of $L\int\Delta\rho$. 
For $W>6$ the collapse of the curves is not clear.}
	\label{DDOS_E_Wall}
\end{figure}

The integrated DOSS, $\Delta\rho(E)$, in the energy window $E_{w}$
is shown in Fig.~\ref{DDOS_E_Wall} as function of disorder strength.
Because Eq.$\,$(\ref{dossL})
suggests a $L^{-1}$ scaling of the DOSS, in Fig.~\ref{DDOS_E_Wall} 
we plot the $L$-rescaled data.
The different data for $L \int\Delta\rho = \tilde{p}L^{-2}$
 seem to collapse in a single curve, approximately
linear, suggesting an exponential decay of the number of surface
states. For high disorder, $W>6$, there are strong fluctuations, 
pointing to a negligible edge signal for such high disorder values.
A fit to the $L=200$
data yields the number of surface states scaling as $\tilde{p}\propto L^{2}\exp(-0.44W)$
in the range $W\in[1,6]$. The number of edge states is a monotonically
decreasing function of $W$, in contrast to the chiral off-diagonal
disorder case, where an enhancement of the number of surface states
has been found for small disorder, up to $W\approx1.75$ \cite{prepar}.

In a clean WNL, the drumhead states are labeled by $\bm{k}_{\parallel}$,
which is a good quantum number. The (exponential) decay length into
the bulk depends on $\bm{k}_{\parallel}$ and diverges as $\bm{k}_{\parallel}$
approaches the nodal line. As stated above, any small amount 
of disorder mixes the clean drumhead states of a WNL: the $\bm{k}_{\parallel}$
labeling of the surface states looses its meaning. However, the surface
states projection onto a plane wave state $\bm{k}_{\parallel}$ in
the plane $z=1$ for sublattice $\alpha=1$ can be probed by computing
the local DOS on the state 
\begin{equation}
|\psi\rangle\equiv|z=1,k_{x},k_{y}=0\rangle=\sum_{x,y}e^{ik_{x}x}|x,y,z=1,\alpha=1\rangle\,,\label{pedro_state}
\end{equation}
at energy $E=0$, defined as 
\begin{align}
 & \Delta\rho(z=1,k_{x},k_{y}=0,E=0)\equiv\nonumber \\
 & \sum_{j({\rm open})}|\langle\psi|j\rangle|^{2}\delta(E_{j})-\sum_{j({\rm closed})}|\langle\psi|j\rangle|^{2}\delta(E_{j})\label{eq:LDOSpedro_state}
\end{align}
where $j$ runs over all quantum eigenstates
of the open or closed system.
The state in Eq.~\eqref{pedro_state} is localized at the $z=1$ plane and therefore the LDOS in it should
be large for surface states.
\begin{figure}[htp]
\centering{}\includegraphics[width=0.95\columnwidth]{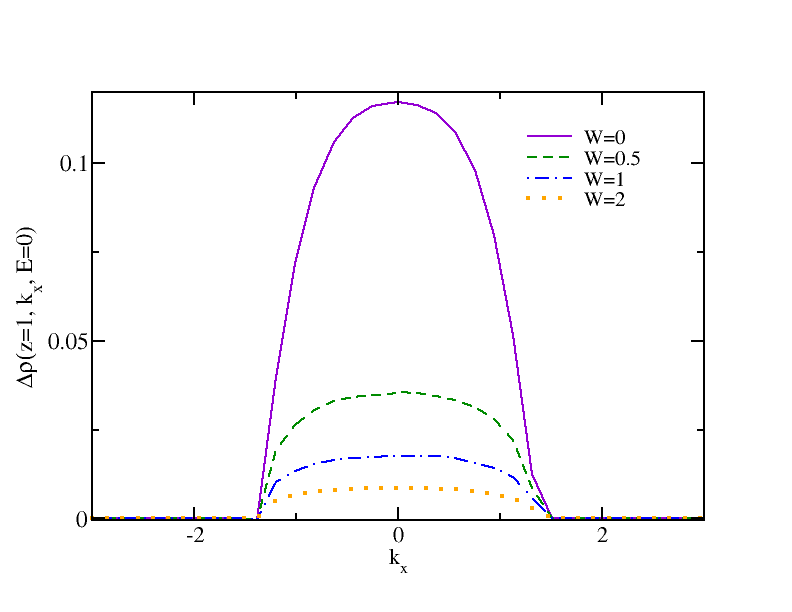}
\caption{LDOS on the state defined in Eq.$\,$(\ref{pedro_state}) for $L=100$,
averaged over 40 disorder realizations.}
	\label{LDOS0_kx_z1}
\end{figure}

The result for the LDOS in state~(\ref{pedro_state}) is plotted in  Fig.~\ref{LDOS0_kx_z1}.
It implies that the surface states contain high weight on the state
in Eq.$\,$(\ref{pedro_state}) in the region inside the nodal loop.
The corresponding quantity for the clean system is shown for comparison.
Clearly, a reminiscence of the nodal line remains for $W=1$. We note
that the KPM calculated DOS value at $E=0$ depends on the bandwidth,
which increases with $W$. This implies that the energy resolution
of the calculation is different for different $W$
\cite{Joao2020}.
Therefore, only the widths of the curves in figure \ref{LDOS0_kx_z1}
can be accurately compared, not their absolute values.

\subsubsection{Localization properties}
\label{subsec:Localization-properties}

\begin{figure}[htp]
	\centering{}\includegraphics[width=0.95\columnwidth]{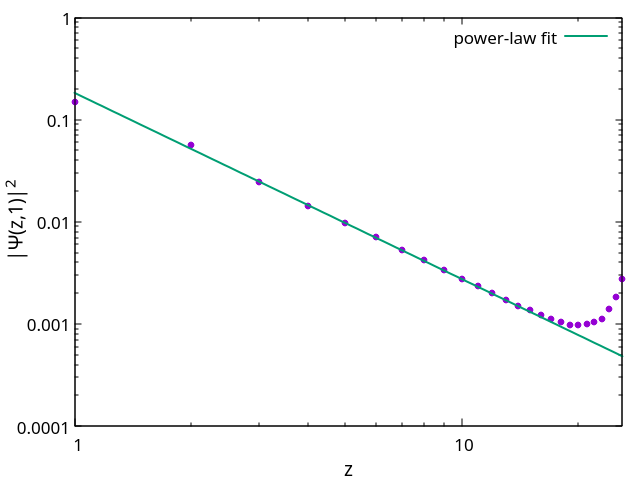}
\caption{Log-log plot of eigenstate's probability projected onto sublattice
$\alpha=1$, for the state
with energy closest to $E=0$, for $L=26$ and $W=0.5$. 
	In this example, the probability decays approximately as $z^{-1.8}$ into
	the bulk.}
\label{psizW05}
\end{figure}

Because of disorder mixing, one can simulate the resulting surface
states as linear superpositions of the clean system's drumhead states
with random coefficients. This yields states that decay algebraically
into the bulk. A realistic
calculation of the surface wave function confirms this expectation.
We used Lanczos ED to find a number of low energy states and calculate the surface
state probability along $z$, as defined in Eq.~(\ref{probz}),
projected on sublattice $\alpha=1$. It is seen that the probability
indeed decays as a power law into the bulk, even for small disorder.
An example is shown in Fig.~\ref{psizW05} for $W=0.5$.
A fit to $\Psi^{2}(z,\alpha)\sim z^{-\nu}$ gives $\nu=1.8$, which
implies an integrable probability as expected for localized states.
Hybridization with the surface state localized in the opposite boundary
is seen for $z>20$. Similar behavior is found for the other low
energy states. By projecting on sublattice $\alpha=2$, we find similar
decaying states from the surface $z=L$, as expected. 
We note that the exponent $\nu$ depends on the hopping parameters in Eq.(\ref{eq:total_H}).
It fluctuates with disorder realization and 
tends to decrease as $W$ increases.

\begin{figure}[htp]
\centering{}\includegraphics[width=0.5\textwidth]{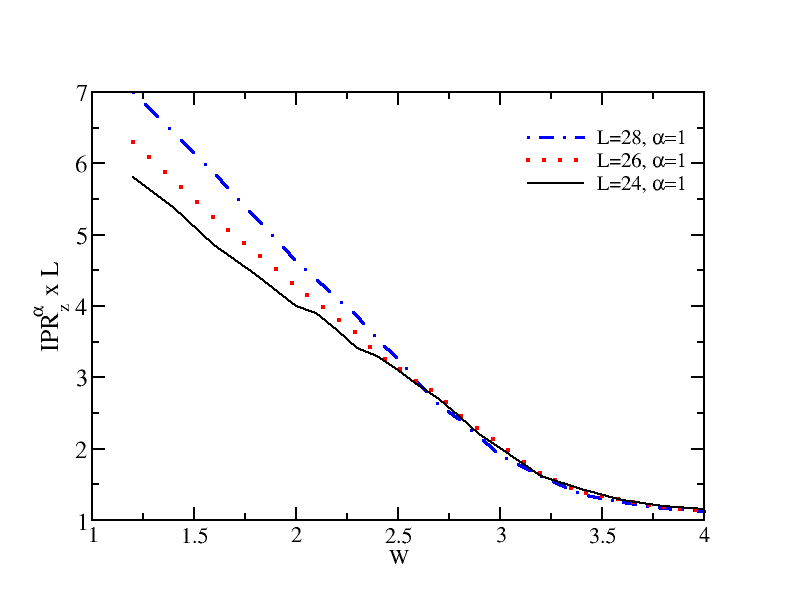}
\caption{$L\times{\rm IPR}_{z}^{\alpha}$ for $L=24\text{\textendash}28$
and the two eigenstates closer to $E=0$, averaged over 400 
disorder realizations. }
	\label{iprzalfa}
\end{figure}

One of the basic principles of the theory of Anderson localization
is that there cannot be two different localization lengths at the
same energy. Since a WNL semimetal has $\rho_{{\rm bulk}}(E)\propto|E|$,
 at finite energy the drumhead states must  hybridize with the extended bulk states and become 
 \textit{delocalized}.
Therefore, a low energy surface state should have the following properties: 
retain high probability near the surface, show power-law decay into the bulk, and 
become extended further into the bulk. At strictly zero energy,
there are no bulk states to hybridize with. This is valid up to the
amount of disorder that renders $\rho_{{\rm bulk}}(0)$ finite, $W\gtrsim2.6$
\cite{Goncalves2020}. Then,
a zero energy ``surface'' state becomes fully extended. 
Evidence for this behavior is shown in Fig.~\ref{iprzalfa}.
It shows ${\rm IPR}_{z}^{\alpha}$ for sublattice $\alpha=1$, as
defined in Eq.~(\ref{iprzalfadef}),
which allows us to study the wave function localization along the
$z$ axis. For extended states in one dimension, the inverse participation
ratio scales with the inverse of the system's size, $L^{-1}$. Such
a scaling is observed in Fig.~\ref{iprzalfa},
where results for different system sizes are plotted. The $L$-rescaled
data collapse into a single curve for  
$W\gtrsim2.6$, thus corroborating the above picture.

\section{Discussion}
\label{sec:Discussion}

We have explored the fate of topological drumhead surface states due
to disorder of the Anderson type in the WNL semimetal.
Because of chiral symmetry breaking, a quantized topological index
(winding number) providing the number of edge states does not exist,
as opposed to the clean case. 
However, in the presence of disorder, a finite sample along $z$
-- the direction perpendicular to the loop plane -- contains extra
states close to zero energy, as compared to the infinite system. Such
states have high probability close to the surface and decay algebraically
into the bulk. Since there is no chiral symmetry, these states can
occur at finite energies ($E\neq0$) due to disorder broadening, and
extend into the bulk because of hybridization with bulk states, which
have finite bulk DOS in the semimetalic phase. 
A similarity can be drawn, here, to the
\textit{virtual bound state} concept in the 
non-interacting Anderson impurity problem, where an initially localized impurity state 
hybridizes with the bulk Bloch waves, thereby producing an extended state
with high probability close to the impurity. 
A concomitant $\pi/2$ scattering phase shift
producing resonant scattering, and 
a broadened DOS peak near the Fermi level appear \cite{Hewson,Doniach}. 
Making an analogy to our case, it is the surface with its exponentially decaying drumhead states
hybridizing with the bulk states that produces a broadened DOS peak near the Fermi level.
Analogously,
the clean system's bulk edge correspondence goes over into
 resonant scattering by the surface in the dirty system.  

At high enough disorder, above the semimetal to metal transition found in Ref.~\cite{Goncalves2020},
even the $E=0$ surface states become extended 
because the bulk DOS, $\rho_{{\rm bulk}}(0)$\lyxadded{Eduardo Castro}{Fri Aug 12 23:10:45 2022}{,}
becomes finite with bulk metallic states. Surface creation brings
about an exponentially decreasing number of extended low energy states,
up to very high disorder. The number of such states is also proportional
to the area of the sample. This is the leftover of the clean WNL semimetal
topology.

We now discuss the similarities and differences to 
recent studies of disorder effects on the Fermi arc states
in nodal point Weyl semimentals (NPWSs) \cite{Slager2017, wilson2018,fedorenko2021}.
In contrast to the drumhead states in a WNL semimental, 
the topological surface states in NPWSs have an energy dispersion which vanishes along a line
(the Fermi arc) on the surface BZ. 
Disorder flattens this energy dispersion \cite{Slager2017} and
enhances the local DOS \cite{fedorenko2021} near the surface.
The finite energy states delocalize because of hybridization with bulk states.
This is similar to the WNL problem, except that the latter's finite energy surface states
emerge from the energy splitting of the initially degenerate clean drumhead states. 
The states on the Fermi arc in a NPWS change from exponentially to 
algebraically
localized,  
and this effect is attributed to hybridization with bulk quasi-localized
states from rare regions \cite{wilson2018}. 
In the WNL case, we attribute the algebraic decay to 
hybridization among clean drumhead states, even without rare region effects.
The complete dissolution of zero energy surface states into the bulk of both 
WNLs and NPWS \cite{fedorenko2021} 
occurs at a finite disorder strength.

Some materials have been confirmed to host nodal lines \cite{HuJin,Jin2017,Ronghan,Liu2018,Sims2019,Schoop2016,Nakamura2019}.
ARPES and transport properties have been used to probe these
materials and also nodal line semimetal candidates \cite{Emma2017,observation_drumhead}. 
An interesting direction for future study is to understand the signatures of the
disorder-driven power-law decaying surface states here unveiled in
ARPES and transport. 
\begin{acknowledgments}
	The authors MA, MG and PR acknowledge partial support from 
Funda\c{c}\~ao para a Ci\^encia e Tecnologia (Portugal) through Grant No. UID/CTM/04540/2019.
JS and EVC acknowledge partial support from 
Funda\c{c}\~ao para a Ci\^encia e
Tecnologia (FCT-Portugal) through Grant No. UIDB/04650/2020. MG acknowledges
further support from FCT-Portugal through the Grant SFRH/BD/145152/2019. 
\end{acknowledgments}

\bibliographystyle{apsrev4-1}
\bibliography{anderson_paper}

\end{document}